\documentclass[aps,pre,preprint,groupedaddress]{revtex4}
\usepackage{graphicx}

\begin{document}


\title{The role of fluctuations in the response of coupled bistable units to
weak driving time-periodic forces}


\author{ Manuel Morillo, Jos\'e G\'omez Ord\'o\~nez and Jos\'e M. Casado}
\email{casado@us.es}
\affiliation{Universidad de Sevilla. Facultad de F\'{\i}sica. \'Area de
F\'{\i}sica Te\'orica. Apartado de Correos 1065. Sevilla 41080. Spain}


\date{\today}

\begin{abstract}
  We analyze the stochastic response of a finite set of globally coupled
  noisy bistable units driven by rather weak time-periodic forces. We
  focus on the stochastic resonance and phase frequency synchronization
  of the collective variable, defined as the arithmetic mean of the variable
  characterizing each element of the array. For single unit systems,
  stochastic resonance can be understood with the powerful tools of
  linear response theory. Proper noise induced phase frequency
  synchronization for a single unit system in this linear response
  regime does not exist. For coupled arrays, our numerical simulations
  indicate an enhancement of the stochastic resonance effects leading to
  gains larger than unity as well as genuine phase frequency
  synchronization. The non-monotonicity of the response with the
  strength of the coupling strength is investigated. Comparison with
  simplifying schemes proposed in the literature to describe the random
  response of the collective variable is carried out.
\end{abstract}


\maketitle


\section{\label{into} Introduction}
The response of noisy nonlinear systems driven by weak external
driving terms is frequently analyzed with the powerful tools of
linear response theory (LRT)\cite{kubo57,hantho82}. The amplification  of the
amplitude of the average response with the noise
strength as well as the non-monotonic behaviors of the amplitude and of the
output signal-to-noise ratio are manifestations of stochastic resonance (SR)
\cite{rmp}. These
behaviors have been indeed rationalized using the ideas of LRT \cite{dyk90}.
Within the linear regime
described by LRT, the response of the non-linear system is
necessarily very noisy. The output noise level is much higher and
the decay time of the fluctuations much longer than those present in
a linear system subject to the same noise and driving terms.
Consequently, the signal-to-noise ratio of the nonlinear system in
that regime is very small, even at its peak value, and the gain does
not exceed unity \cite{casden03}. If one is interested in obtaining large
average
amplitudes and large signal-to-noise ratio, one must find a way to
control the output fluctuations and this is only possible in a
nonlinear regime.

A possible way to reduce the output fluctuation levels in noise induced
SR is to use external driving terms with amplitudes so large that, even
though they are still subthreshold, they alter significantly the
potential relief of the dynamics and render invalid the LRT assumptions
\cite{casgom03}. Another possibility is to concentrate on the global
response of a set of nonlinear oscillators. In
\cite{jung,schm,our,neiman} the enhancement of SR effects in arrays of
bistable systems were studied.

More recently \cite{us06,allus07}, we have analyzed the collective
response of a finite set of globally coupled bistable systems. We
demonstrated that SR is indeed much enhanced with respect to SR in
single bistable units. Indeed, gains larger than unity were observed for
subthreshold sinusoidal driving forces. Those findings indicated that
the arrays were indeed operating in nonlinear regimes.

Another aspect of the stochastic response of the system refers to
its noise induced synchronization with the driving term. By a
suitable definition of a phase associated to the random output
process, one can define an output average phase frequency and a
phase dispersion. The matching of the phase frequency with the
driving frequency for a range of noise values is what is termed noise
induced frequency synchronization. It has been analyzed with a
variety of analytical and numerical procedures
\cite{pikovskybook2001, chaos2003, vainstein1983, schimansky2001},
even in circumstances where quantum tunneling effects are relevant
\cite{goychuk}. In \cite{us07} noise induced phase synchronization
of the collective variable for wide ranges of noise values were also
observed when the array was driven by subthreshold inputs.

The results reported in the above mentioned work were obtained with
driving terms that, even though they were unable to produce the
cited effects in the absence of noise, they were large enough to
invalidate LRT when applied to a single bistable system. The
question we address in the present work is whether the enhancement
of SR and synchronization effects in the collective response of
finite arrays of bistable units still persist when the driving
external force is rather weak. By weak, we will mean here that: i)
the SR effects induced by the external driving in a single bistable
unit can be satisfactorily described, even at a quantitative level,
by the LRT approximation; and ii) noise induced phase frequency
synchronization in a single bistable unit does not properly exist.

In the next Section, we introduce the model system and define the
relevant quantities that characterize the phenomenon of SR and phase
synchronization. In Section \ref{results} we show the results
obtained by numerically solving the dynamical stochastic equations.
Comparison with the predictions of recently formulated approximate
descriptions of the collective dynamics \cite{Pikovsky,cubero} is
carried out in Section IV. The last section concludes with some
remarks.

\section{\label{model} Model and definitions}

We consider a set of $N$ identical subsystems, each of them
characterized by a variable $x_i(t)\, (i=1,\ldots,N)$ satisfying a
stochastic evolution equation (in dimensionless form) of the type
\cite{deszwa,casgom06}
\begin{equation}
\dot{x}_i=x_i-x_i^3+\frac{\theta}{N}\sum_{j=1}^N(x_j-x_i)+\sqrt{2D}\xi_i(t)+F(t),
\label{eq:lang}
\end{equation}
The external driving force is periodic in $t$, $F(t)=F(t+T)$.  The
term $\xi_i(t)$ represents a white noise with zero average and
$\left\langle \xi_i(t) \xi_j(s) \right\rangle = \delta_{ij}\delta
(t-s)$. We define a collective variable, $S(t)$,
\begin{equation}
S(t)=\frac 1N \sum_j x_j(t),
\end{equation}
and concentrate on its long time response when the system size, $N$,
is kept finite and the amplitude of the driving term is weak (in the
sense that when the driving force acts on a single isolated unit, SR
is well described by linear response theory). We define the one-time
correlation function,
\begin{equation}
L(\tau)=\frac 1T \int_0^T \,dt\; \langle S(t)S(t+\tau)\rangle_{*} ,
\end{equation}
The notation $\langle \ldots \rangle$ indicates an average over the
noise realizations and the subindex $*$ indicates the long time
limit of the noise average, i. e., its value after waiting for $t$
large enough that transients have died out. As shown in our previous
work, \cite{casgom06} we have that
\begin{equation}
L(\tau)=L_\mathrm{coh}(\tau)+L_\mathrm{incoh}(\tau)
\end{equation}
where the coherent part, $L_{coh}(\tau)$,  is periodic in $\tau$
with the period of the driving force, while the incoherent part,
$L_{incoh}(\tau)$ arising from the fluctuations of the output $S(t)$
around its average value, decays to zero as $\tau$ increases. The
output signal-to-noise ratio, $R_{out}$ is
\begin{equation}
\label{snr} R_\mathrm{out} =\lim_{\epsilon \rightarrow 0^+}\frac {
\int_{\Omega-\epsilon}^{\Omega+\epsilon} d\omega\;
\tilde{L}(\omega)}{\tilde{L}_\mathrm{incoh}(\Omega)}=\frac {
\tilde{L}_\mathrm{coh}(\Omega)}{\tilde{L}_\mathrm{incoh}(\Omega)} ,
\end{equation}
where $\Omega $ is the fundamental frequency of the driving force
$F(t)$, $\tilde{L}_\mathrm{coh}(\Omega)$ is the corresponding
Fourier coefficient in the Fourier series expansion of
$L_\mathrm{coh}(\tau)$, and $\tilde{L}_\mathrm{incoh}(\Omega)$ is
the Fourier transform at frequency $\Omega$ of
$L_\mathrm{incoh}(\tau)$.

For a set of $N$ coupled linear oscillators driven by an external
driving force $F(t)$ and subject to the noise terms $\xi_i(t)$ as in
Eq. (\ref{eq:lang}), the SNR of the corresponding collective
process, $R_\mathrm{out}^{(L)}$, coincides with that of the random
process formed by the arithmetic mean of the individual noise terms
$\xi_i(t)$ plus the deterministic driving force $F(t)$, namely,
$F(t)+\xi(t)$ with $\xi(t)=N^{-1}\sum_{i=1}^N\xi_i(t)$.  The process
$\xi(t)$ is a Gaussian white noise of effective strength $D/N$.
Then, it is easy to prove that
\begin{equation}
R_\mathrm{out}^{(L)}=\frac{2 A^2 N [1-\cos(\pi r)]}{\pi D}.
\label{eq:Rin}
\end{equation}
Thus, for our nonlinear case, it seems convenient to analyze the SR
gain, $G$, defined as \cite{casgom06}
\begin{equation}
\label{gain} G=\frac{R_\mathrm{out}}{R_\mathrm{out}^{(L)}},
\end{equation}
which compares the SNR of a non-linear system with that of a linear
system subject to the same stochastic and deterministic forces.

In the case of noninteracting units ($\theta=0$), the SNR of the
collective output is $N$ times larger than that of a isolated unit
driven by the same force. Nonetheless, as discussed in
\cite{casgom06}, the gain associated to the collective output is
just the same as the one of a single, isolated, unit. Thus, for the
weak forces that we are considering here, we expect that the
collective gain will not exceed unity. As seen below, our numerical
results will confirm that result.

Another aspect of the response is the noise induced phase frequency
synchronization. We note that for low noise strengths and driving
forces with sufficiently large periods, a random trajectory of
$S(t)$ contains essentially small fluctuations around two values
(attractors) and random, sporadic transitions between them. For each
realization of the noise term, we then introduce a random phase
process, $\phi(t)$, associated to the stochastic variable $S(t)$ as
follows.  We refer to a ``jump'' of $S(t)$ along a trajectory, when
a very large fluctuation takes the $S(t)$ trajectory from a value
near an attractor to a value in the neighborhood of the other
attractor.  We count $N^{(\alpha)}(t)$, the number of jumps in the
$\alpha$ trajectory of the process $S(t)$ within the interval
$(0,t]$ . A trajectory of the phase process is then constructed as
\begin{equation}
\label{phase}
\phi^{(\alpha)}(t)=\pi N^{(\alpha)}(t),
\end{equation}
so that $\phi(t)$ increases by $2\pi$ after every two consecutive
jumps.

 The first two moments of the phase process are estimated as
\begin{equation}
\label{phaseaverage} \left\langle \phi(t)\right\rangle
=\frac{1}{\mathcal M} \sum_{\alpha=1}^{\mathcal M}
\phi^{(\alpha)}(t)
\end{equation}
\begin{eqnarray}
\label{phasedispersion} v(t)&=&\left\langle
\left[\phi(t)\right]^2\right\rangle -\left\langle
\phi(t)\right\rangle^2 \nonumber \\
&=&\frac{1}{\mathcal M} \sum_{\alpha=1}^{\mathcal M}
\left[\phi^{(\alpha)}(t)\right]^2-\frac{1}{\mathcal M^2}
\left[\sum_{\alpha=1}^{\mathcal M} \phi^{(\alpha)}(t)\right]^2
\end{eqnarray}
where, ${\mathcal M}$ is the number of generated random
trajectories.

The instantaneous phase frequency is easily determined from the time
derivative of $\left\langle \phi(t)\right\rangle $.  After a
sufficiently large number of periods of the driving force, $n$, the
system forgets its initial preparation, but the instantaneous phase
frequency is still a function of time. Then, we define a cycle
average phase frequency $\overline{\Omega}_\mathrm{ph}$ by averaging
the instantaneous phase frequency over a period of the external
driving, \cite {casado,goychuk}

\begin{equation}
\label{frequency2} \overline{\Omega}_\mathrm{ph}=\frac 1T
\int_{nT}^{(n+1)T} dt\,\frac {d\langle \phi(t) \rangle}{dt}=
\frac{\left\langle \phi\left[(n+1)T\right]\right\rangle-\left\langle
\phi(nT)\right\rangle}{T}
\end{equation}

Similarly, the cycle average phase diffusion coefficient is
evaluated from the instantaneous slope of the variance $v(t)$ as
\cite{casado,goychuk},
\begin{equation}
\label{diffusion2} \overline{D}_\mathrm{ph}=\frac 1T
\int_{nT}^{(n+1)T} dt\,\frac {d\langle v(t) \rangle}{dt}=
\frac{v\left[(n+1)T\right]
  -v\left(nT\right)}{T}
\end{equation}
In previous works, approximate analytical expressions for these two
quantities have been derived for the $N=1$ problem in the classical
\cite{freund1,casado,talkner1,prager} and quantum cases
\cite{goychuk}. Those expressions can not be applied to the
collective variable of an $N$-particle problem,

\section{\label{results}Numerical results}
In general, nonlinearities preclude exact analytical solutions of
Eqs.\ (\ref{eq:lang}). We will use numerical simulations to obtain
useful information about the stochastic process $S(t)$. In the
asymptotic limit $N\rightarrow \infty$, Desai and Zwanzig showed
that the statistical properties of the model could be analyzed in
terms of a nonlinear Fokker-Planck equation which allows the
coexistence of several stable probability distributions for some
ranges of noise strengths and $\theta < 1$.  In the same asymptotic
limit, we analyzed a few years ago the stochastic resonant behavior
of the first moment, $\langle S(t) \rangle_*$, when the system is
driven by a time dependent sinusoidal force, using a combination of
analytical and numerical procedures \cite{our}. In particular, for
very weak input amplitudes, a linear response theory analysis showed
that a huge amplification in the amplitude of the average output
$\langle S(t) \rangle_*$ with respect to that of the driving force
could be achieved.

In this work we concentrate on situations where: i) the number of
subunits is finite; ii) the amplitude of the driving force is rather
weak. We have considered  an external periodic rectangular driving,
\begin{equation}
\label{force}
F(t)=(-1)^{n(t)} A,
\label{Eq009}
\end{equation}
where $n(t)=\lfloor 2\, t/T \rfloor$, $\lfloor z \rfloor$ is the
floor function of $z$, i.e., the greatest integer less than or equal
to $z$. In other words, $F(t)=A$ ($F(t)=-A$) if $t\in [n T/2,
(n+1)T/2)$ with $n$ even (odd). As detailed in \cite{casgom03} the
numerical algorithm used to integrate the Langevin equations follows
one of the schemes put forward by Greenside and Helfand
\cite{greenhel}.

\begin{figure}
\includegraphics[width=8cm]{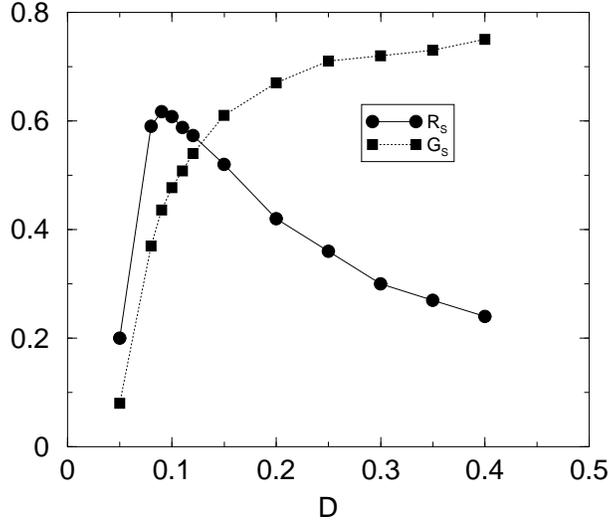}%
\caption{\label{uno} Signal-to-noise ratio, $R_S$ (circles) and gain
  $G_S$ (squares) of the collective variable for an array of $N=10$
  non-coupled, ($\theta=0$), bistable units driven by rectangular inputs
  with $\Omega=0.01$ and $A=0.1$.}
\end{figure}

\begin{figure}
\includegraphics[width=8cm]{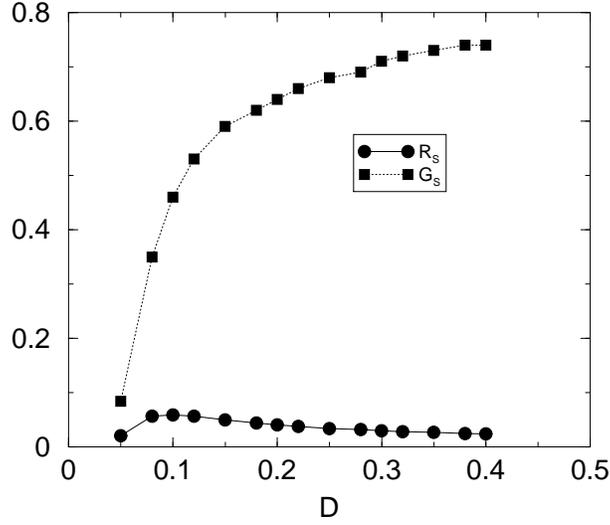}
\caption{\label{dos} Signal-to-noise ratio $R_S$ (circles) and gain
  $G_S$ (squares) for a single bistable unit driven by rectangular
  inputs with $\Omega=0.01$ and $A=0.1$.}
\end{figure}

Let us first consider the case of independent subunits, $\theta=0$.
In Fig. \ref{uno} we display the behavior with respect to the noise
strength, $D$, of the signal-to-noise ratio and the gain of the
output variable $S(t)$ for an array of 10 noninteracting bistable
units, driven by a weak amplitude rectangular force ($A=0.1$) with
fundamental frequency $\Omega=0.01$. For comparison purposes, in
Fig. \ref{dos} we present the results for the signal-to-noise ratio
and the gain of the response of a single bistable units operating
under the action of the same driving term as in the previous figure
Clearly, SR is manifested in the non-monotonic behavior of the
signal-to-noise ratio with respect to the noise strength in both
figures. But, as expected, the $R$ values for the $N=1$ case are
very small, $1/10$ times the $R$ values for the $N=10$ case. On the
other hand, the gain in both cases has the same values, not
exceeding unity as required by the linear response theory. The
expected amplification of the $R$ values are in agreement with the
predictions of the central limit theorem which indicates a reduction
by an $1/N$ factor of the output fluctuations of the array with
respect to those of a system with a single unit.

\begin{figure}
\includegraphics[width=8cm]{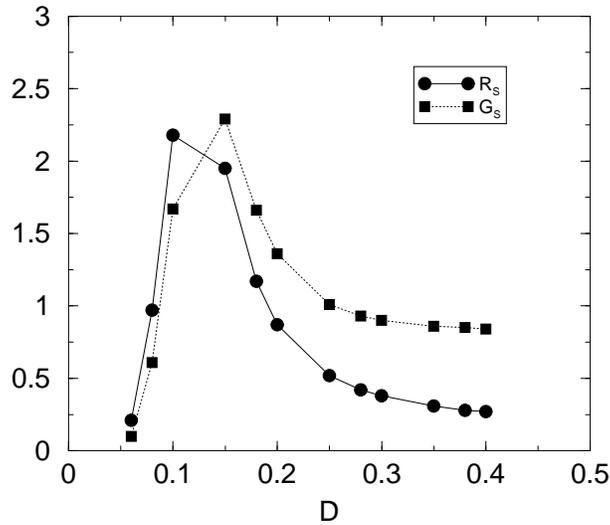}%
\caption{\label{tres} Signal-to-noise ratio $R_S$ (circles) and gain
  $G_S$ (squares) of the
collective variable for an array of $N=10$ bistable units driven by
rectangular inputs
  with $\Omega=0.01$ and $A=0.1$ and coupling strength $\theta=0.2$.}
\end{figure}

The interactions between the units bring up changes in the
collective response as depicted in Fig. \ref{tres}. Here we have
coupled the $N=10$ bistable elements with a weak coupling strength
$\theta=0.2$. The existence of coupling increases substantially the
non-monotonic behavior of $R$ vs. $D$ of the collective output
relative to the uncoupled units case in Fig. \ref{uno}. Also, the
peak value is reached at higher values of $D$. And, more
importantly, the gain is clearly above unity for a wide range of
noise values. This fact indicates that the array is operating in a
nonlinear regime even though the driving amplitude $A$ is rather
small.

The dependence on the interaction strength $\theta$ of $R$ and $G$
is explicitly demonstrated in the next figure Fig. \ref{cuatro}.
Here, we depict the peak values of the signal-to-noise ratio,
$R_\mathrm{max}$, and the gain $G_\mathrm{max}$ on the coupling
strength $\theta$ for an array of $N=10$ elements driven by a
rectangular input with parameters $\Omega=0.01$ and $A=0.1$. The
non-monotonic behavior with $\theta$ is clear. As the coupling
strength increases from zero the SR effects become more pronounced
until they reach a maximum at around $\theta=0.5$. Increasing
further the coupling strength leads to a decrease in the peak
values.

\begin{figure}
\includegraphics[width=8cm]{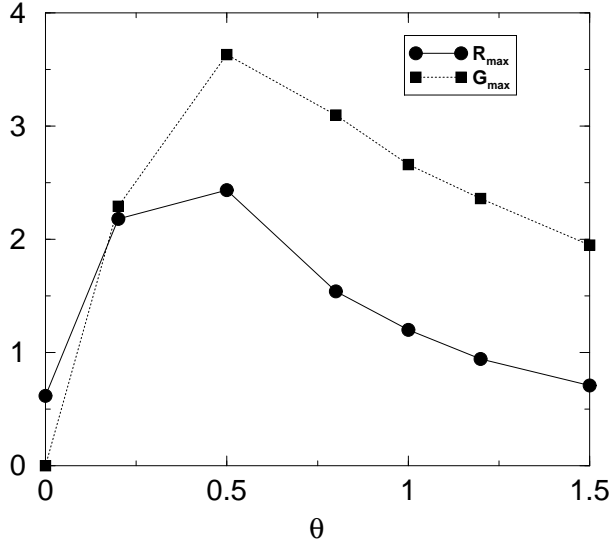}%
\caption{\label{cuatro} The peak values of the signal-to-noise ratio
($R_\mathrm{max}$) and gain ($G_\mathrm{max}$) of a $N=10$ array of
bistable units vs. the coupling strength $\theta$. The driving term
is  rectangular
  with $\Omega=0.01$ and $A=0.1$.}
\end{figure}

\begin{figure}
\includegraphics[width=8cm]{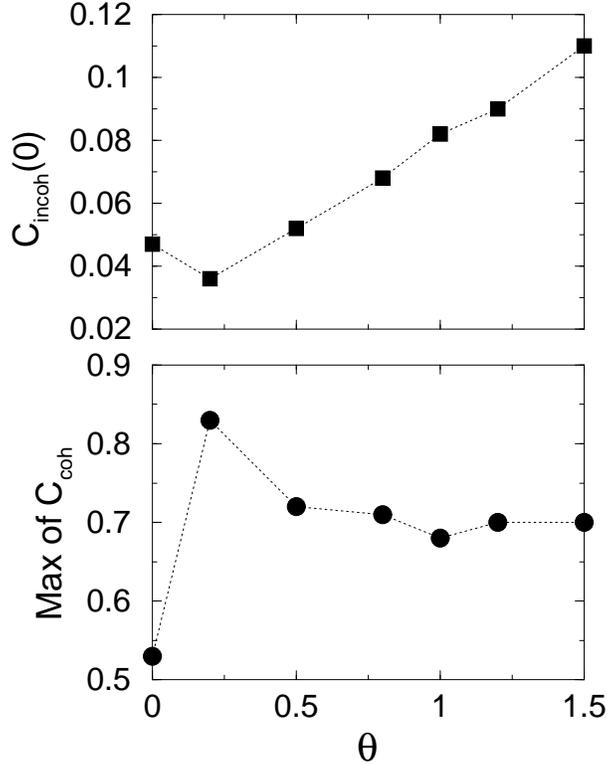}%
\caption{\label{cinco} The maximum value of $C_\mathrm{coh}(t)$ and the
  initial value of the incoherent part,  $C_\mathrm{incoh}(0)$ for
  several values of $\theta$}
\end{figure}

The signal-to-noise ratio depends on the behavior of the correlation
function. In the lower panel in Fig. \ref{cinco}, we depict the
maximum value of the coherent part of the correlation function
($C_\mathrm{coh}$) on the coupling strength $\theta$ for those noise
values at which the amplification is maximal. In the upper panel,
the dependence of the initial value of the corresponding incoherent
part $C_\mathrm{incoh}(0)$ with $\theta$ is shown. One observes that
both quantities show non-monotonic behaviors with the coupling
strength. Even though the peak in the lower panel and the minimum in
the higher panel are obtained for $\theta=0.2$, the largest
signal-to-noise ratio is reached at the slightly larger coupling
strength $\theta=0.5$. This can be understood by noting that, as
depicted in Fig. \ref{cinco}, the decay of the incoherent part of
the correlation function at $\theta=0.5$ is faster than at the other
coupling values. Indeed, the key behavior is that of the incoherent
part of the correlation function. Comparing $C_\mathrm{incoh}(t)$,
for $\theta=0$ and $\theta=0.2$ we see that their initial values get
smaller as $\theta$ increases, while the decay time of the
correlation function remains practically the same. On the other
hand, as $\theta$ increases to $0.5$, the initial value also
increases but the decay is much faster. Consequently, the
denominator in the ratio of Eq.\ (\ref{snr}) decreases and the ratio
reaches a maximum value. As the coupling constant is increased
further, the initial values increase and therefore the values of $R$
decrease. These features indicate that large enhancements of the SR
quantifiers and large gain values are achieved when the output
fluctuations are small and fast decaying. These two combined facts
can only be achieved when the system operates in a nonlinear regime.
For a weak driving force this nonlinear regime is not possible with
a single unit system or with an array of noninteracting units.

\begin{figure}
\includegraphics[width=8cm]{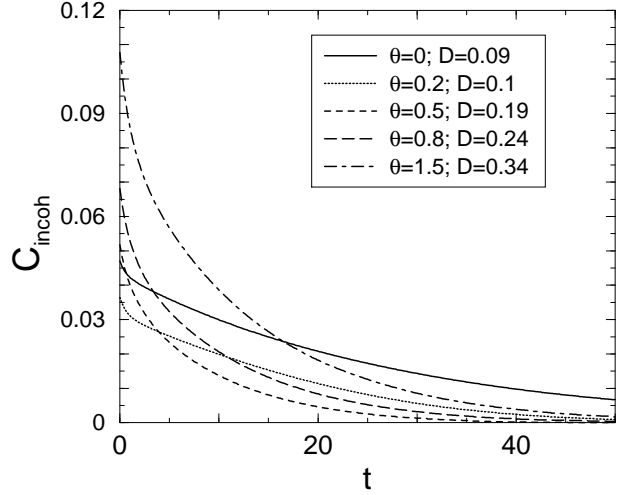}%
\caption{\label{seis} Coherent part of the correlation functions of
the collective variable $S(t)$ for $N=10$ and several values of
$\theta$. The driving term is rectangular
  with $\Omega=0.01$ and $A=0.1$. The noise values used correspond to the
  values at which $R$ reaches its peak for the different coupling
  strength.}
\end{figure}

\begin{figure}
\includegraphics[width=8cm]{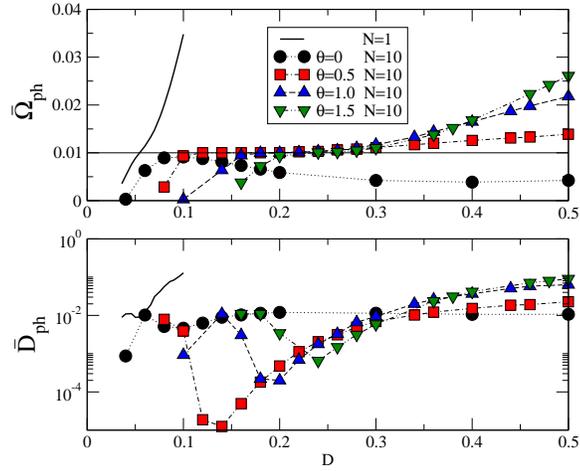}%
\caption{\label{siete} The phase frequency and the phase diffusion coefficient
  as  functions of $D$.
  The driving term is rectangular with $\Omega=0.01$ and $A=0.1$. }
\end{figure}
Let us know turn our attention to the noise induced synchronization.
In Fig.\ \ref{ocho} the phase frequency ($
\overline{\Omega}_\mathrm{ph}$) and the phase diffusion coefficient
( $\overline{D}_\mathrm{ph}$) obtained from numerical simulations
are depicted. The driving force in all cases is a rectangular input
with amplitude $A=0.1$ and fundamental frequency $\Omega=0.01$. It
is clear that for this weak driving force and for a single unit
system, the range of noise values at which the phase frequency
matches the driving fundamental frequency is extremely narrow to
consider that there is a proper synchronization. On the other hand,
for $N=10$ coupled particles, the range of noise values leading to
proper frequency matching and small phase diffusion coefficients is
substantial. Optimum synchronization is obtained for $\theta=0.5$
with a dip at the diffusion constant for $D\approx 0.15$. This noise
value is slightly smaller than the one at which the signal-to-noise
ratio reaches its maximum value for the same coupling constant
($D\approx 0.19$). Note that as the value of $\theta$ increases
above $0.5$, the range of noise values for synchronization to take
place reduces and it shifts to higher values of $D$. Perhaps, the
most surprising result is that for uncoupled units, synchronization
between the collective variable and the driving term is lost. This
is indeed in sharp contrast with the results reported in \cite{us07}
for the same system with a driving force with a much larger
amplitude $A=0.3$.

The range of noise values for stochastic resonance  and noise
induced phase synchronization do not have to coincide. Indeed, our
results indicate that SR might be present at parameter values where
noise induced phase synchronization does not exist. This is not
surprising. Both effects are aspects of the stochastic response of
the system to a driving agent. They are undoubtedly related, but
they probe different aspects of that response.

\section{\label{compare}Comparison with approximate descriptions}
Due to the nonlinearity of the dynamical equations, the collective
variable does not obey a single closed Langevin equation, but an
infinite hierarchy of equations. Recently, for finite arrays,
approximate descriptions of the dynamics based on truncations of the
infinite hierarchy of fluctuating cumulants have been put forward by Pikovsky et
al. \cite{Pikovsky} and by Cubero \cite{cubero}. The Gaussian
approximation amounts to describe the dynamics of the collective
variable by the equations \cite{Pikovsky},

\begin{eqnarray}
\label{gaussian}
  \dot{S}&=& S-S^3-3MS+\sqrt{\frac {2D}N}\xi(t)+F(t) \\
 \frac 12 \dot{M} &=& M-3S^2M-3M^2-\theta M+D.
\end{eqnarray}

 Even within the reduced
Gaussian approximation, one has to rely on numerical treatments to
obtain reliable information. The predictions of the Gaussian
approximation (without invoking a slaving principle) are compared
with those obtained from the full solution of the entire dynamics in
the next figures.

\begin{figure}
\includegraphics[width=8cm]{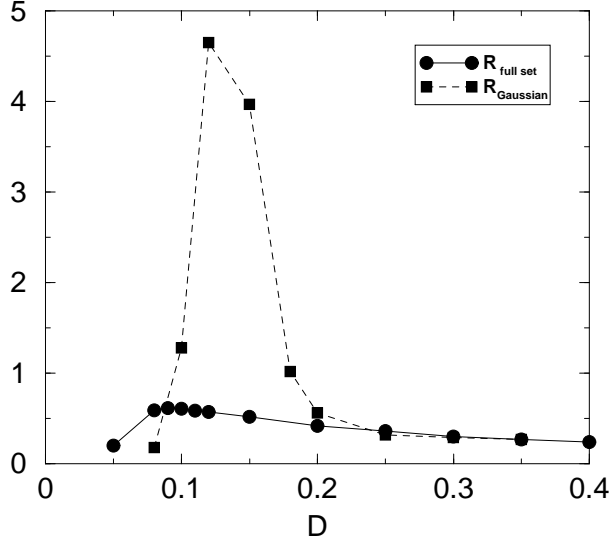}%
\caption{\label{ocho} Comparing the signal-to-noise ratio as obtained
  from the numerical simulation of the full set of equations and from
  the Gaussian approximation for $N=10$ uncoupled units(
  $\theta=0$). Other parameter values: $\Omega=0.01$ and $A=0.1$.}
\end{figure}

\begin{figure}
\includegraphics[width=8cm]{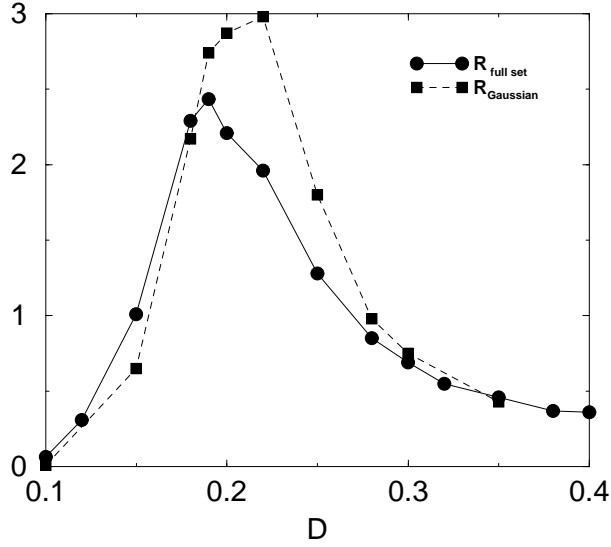}
\caption{\label{nueve} Comparing the signal-to-noise ratio as obtained
  from the numerical simulation of the full set of equations and from
  the Gaussian approximation for $N=10$ weakly ($\theta=0.5$) coupled
  units. Other parameter values: $\Omega=0.01$ and $A=0.1$.}
\end{figure}

\begin{figure}
\includegraphics[width=8cm]{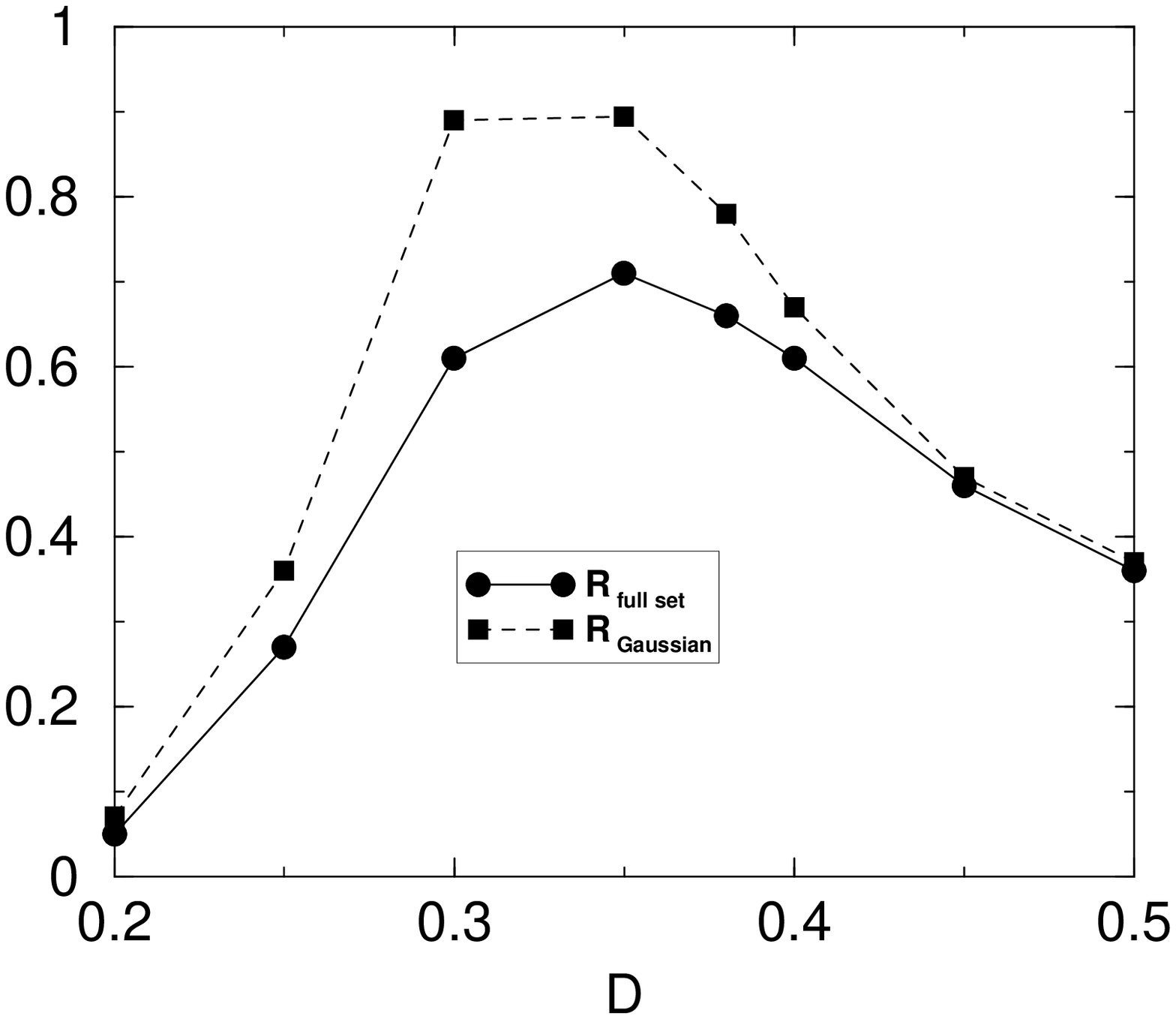}%
\caption{\label{diez}
  Comparing the signal-to-noise ratio as
obtained from the numerical simulation of the full set of equations
and from the Gaussian approximation for $N=10$ coupled units with
coupling strength  $\theta=1.5$. Other parameter
values: $\Omega=0.01$ and $A=0.1$.}
\end{figure}

In Figs.\ \ref{ocho}, \ref{nueve}, \ref{diez} we show the results
obtained for the noise dependence of the collective signal-to-noise
ratio as given by the simulation of the full set of Langevin
equations and by the Gaussian approximation. In Fig.\ \ref{ocho} we
consider the case of uncoupled arrays, while in Figs.\ \ref{nueve}
and \ref{diez} we consider  moderate ($\theta=0.5$) and strong
($\theta=1.5$) coupling cases respectively. The Gaussian
approximation seems to be more reliable as the value of the coupling
strength increases, although it yields poor results around the noise
values at which $R$ shows its peak. Only for high values of the
noise strength does the Gaussian approximation lead to results in
good agreement with those obtained from the full set of equations.

As discussed in \cite{Pikovsky}, the Gaussian truncation combined
with a slaving principle, allows for a simplified description of the
dynamics of $S(t)$ in terms of a Langevin equation in an effective
double well potential and a noise term $\sqrt{\frac {2D}N}\xi(t)$.
This proposal (or the effective potential proposed in
\cite{cubero}), has the attractive feature of reducing the dynamics
for the collective variable to a single Langevin equation in a
one-dimensional potential.  Unfortunately, by contrast with the
parameter values for $\theta$ and $D$ considered in those works, for
the parameter values used here the effective potential evaluated as
indicated in the above mentioned references might not be bounded or
it might not even exist.

\begin{figure}
\includegraphics[width=8cm]{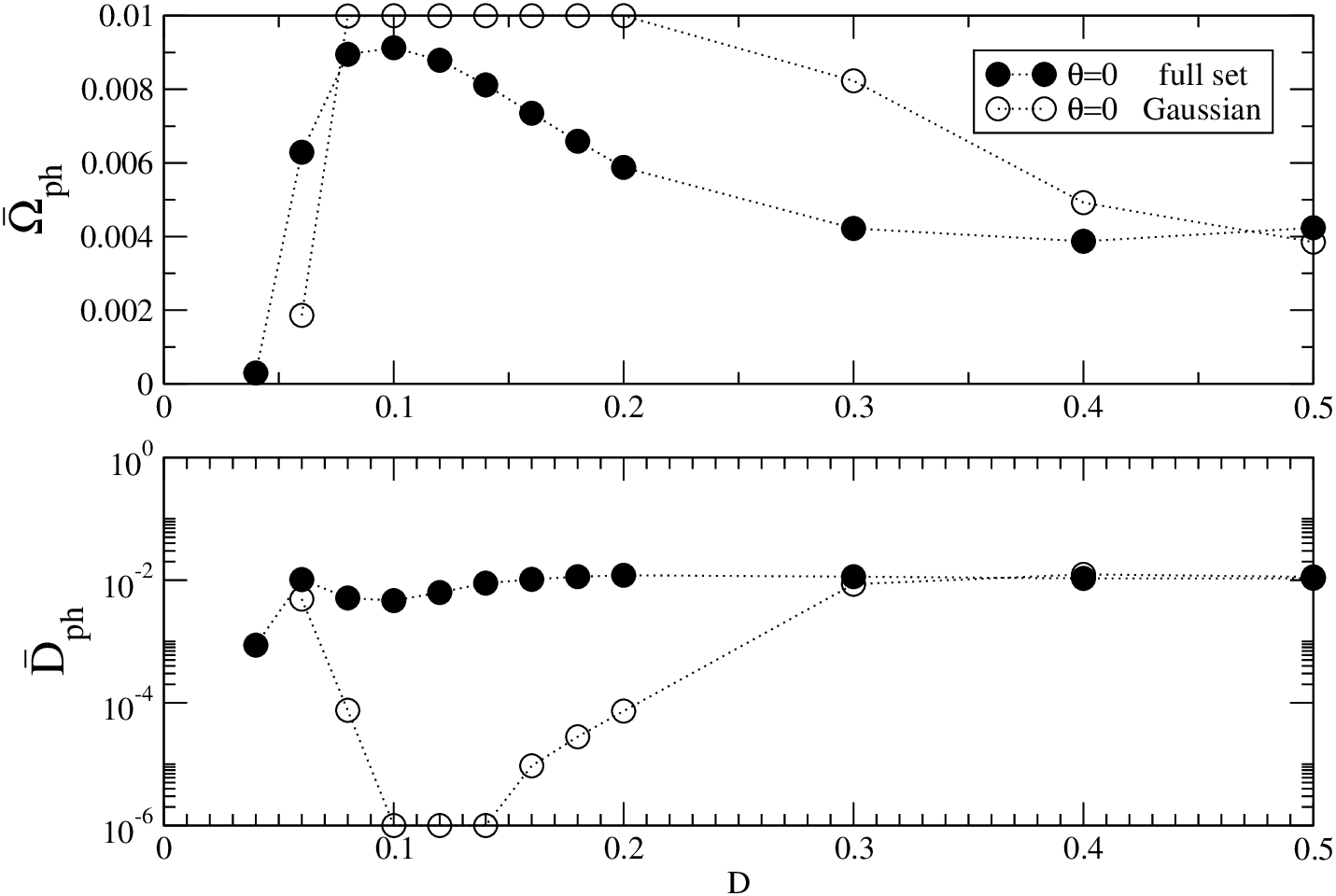}%
\caption{\label{once} Comparing average phase frequency and phase
  diffusion as obtained from the numerical simulation of the full set of
  equations and from the Gaussian approximation for $N=10$ and
  ($\theta=0$). Other parameter values: $\Omega=0.01$
  and $A=0.1$.}
\end{figure}

We have also compared the results obtained with the Gaussian
truncation for the average phase frequency and the phase diffusion
coefficient with respect to the ones obtained from the full set of
Langevin equations. In Fig.\ \ref{once}, we depict the results
obtained for $N=10$ uncoupled units ($\theta=0$) driven by a
rectangular periodic force with  $A=0.1$ and fundamental frequency
$\Omega=0.01$. The Gaussian approximation predicts a perfect
matching of the driving frequency and the phase frequency for a wide
range of noise strength values. These results are completely at
variance with those obtained from the numerical solution of the full
set of equations. Thus, the Gaussian approximation is not reliable
for the $\theta=0$ case.

\begin{figure}
\includegraphics[width=8cm]{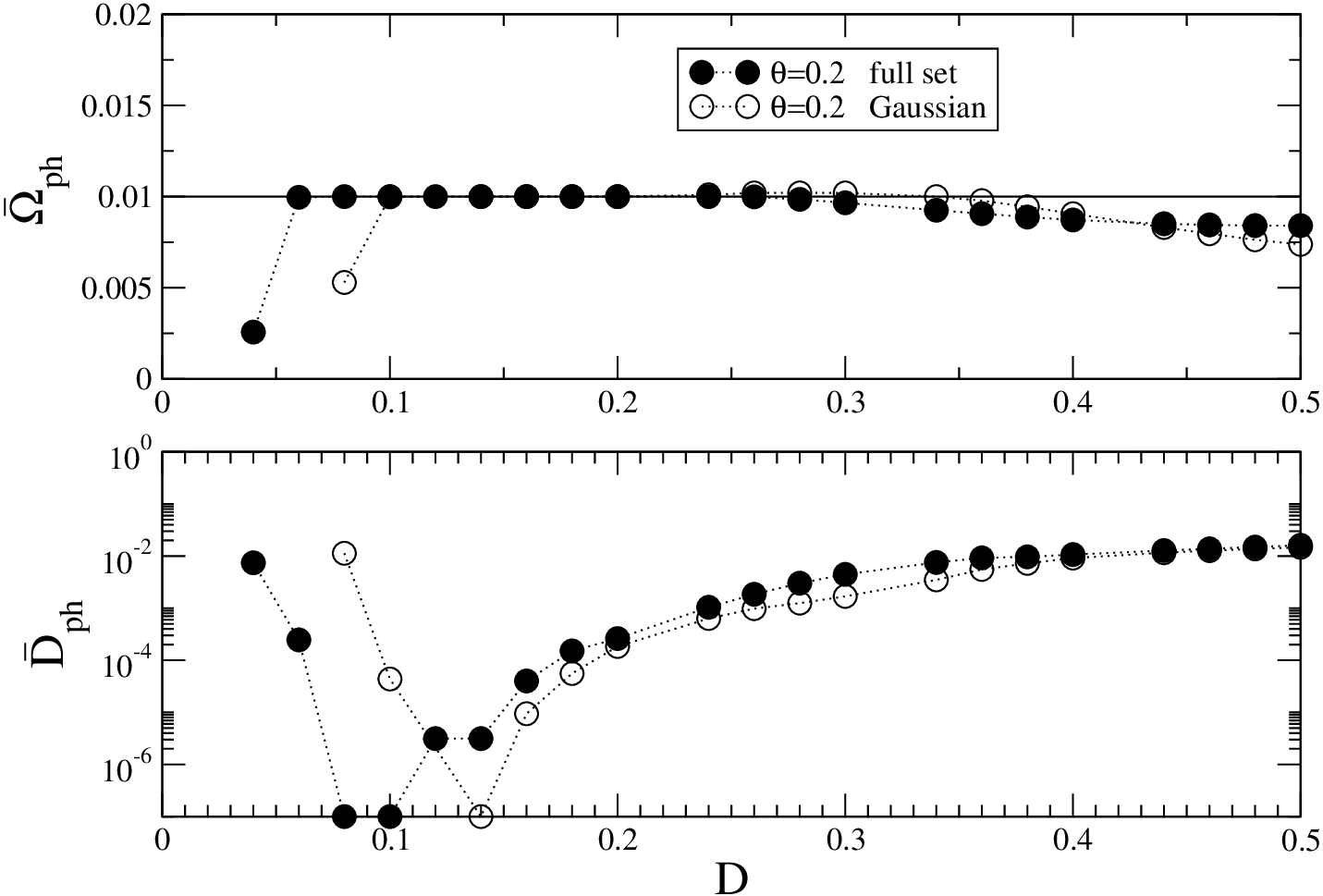}%
\caption{\label{doce} Comparing average phase frequency and phase
  diffusion as obtained from the numerical simulation of the full set of
  equations and from the Gaussian approximation for $N=10$ and
  ($\theta=0.2$). Other parameter values: $\Omega=0.01$
  and $A=0.1$.}
\end{figure}

\begin{figure}
\includegraphics[width=8cm]{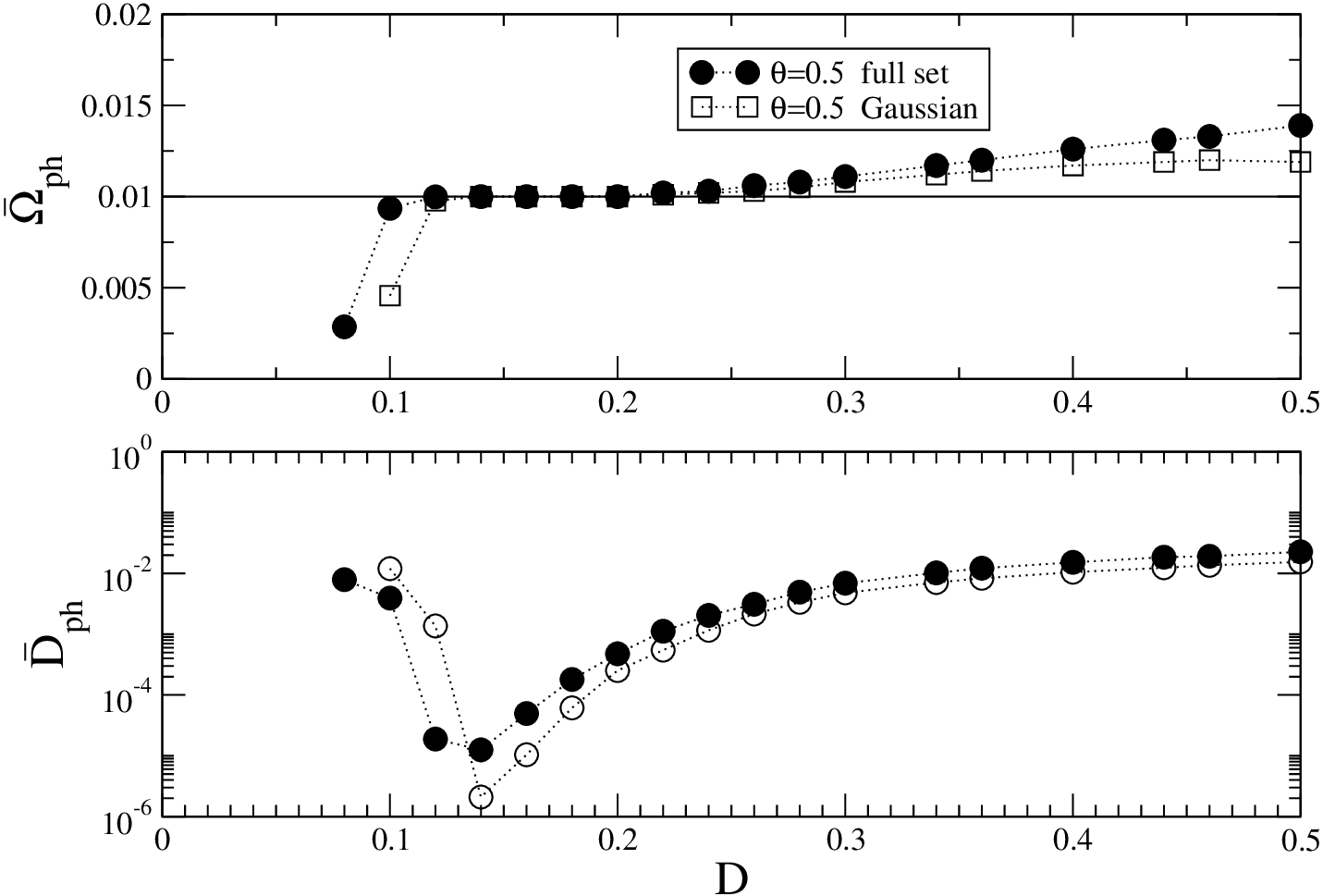}%
\caption{\label{trece} Comparing average phase frequency and phase
  diffusion as obtained from the numerical simulation of the full set of
  equations and from the Gaussian approximation for $N=10$ and
  ($\theta=0.5$). Other parameter values: $\Omega=0.01$
  and $A=0.1$.}
\end{figure}

\begin{figure}
\includegraphics[width=8cm]{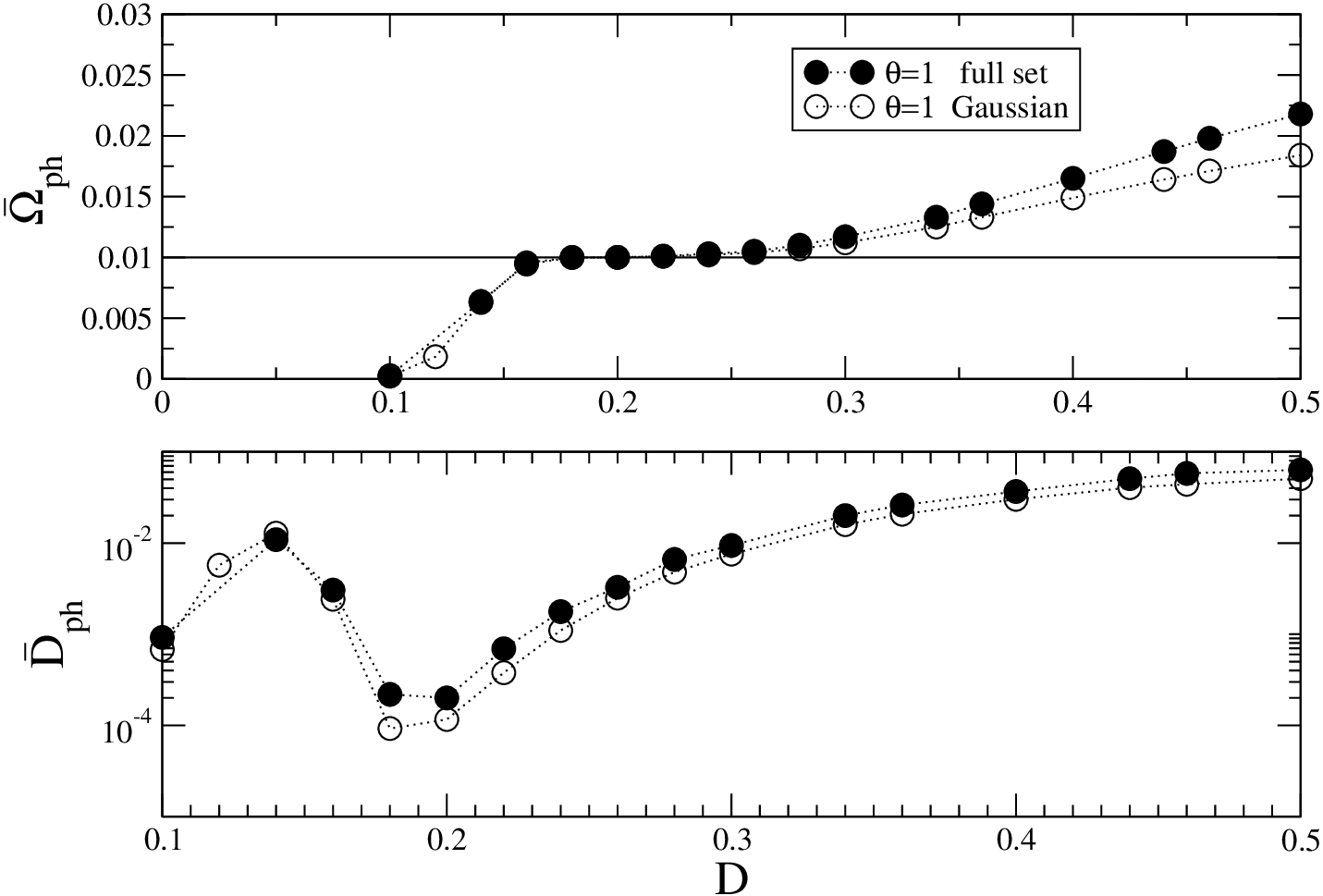}%
\caption{\label{catorce} Comparing average phase frequency and phase
  diffusion as obtained from the numerical simulation of the full set of
  equations and from the Gaussian approximation for $N=10$ and
  ($\theta=1$). Other parameter values: $\Omega=0.01$
  and $A=0.1$.}
\end{figure}

\begin{figure}
\includegraphics[width=8cm]{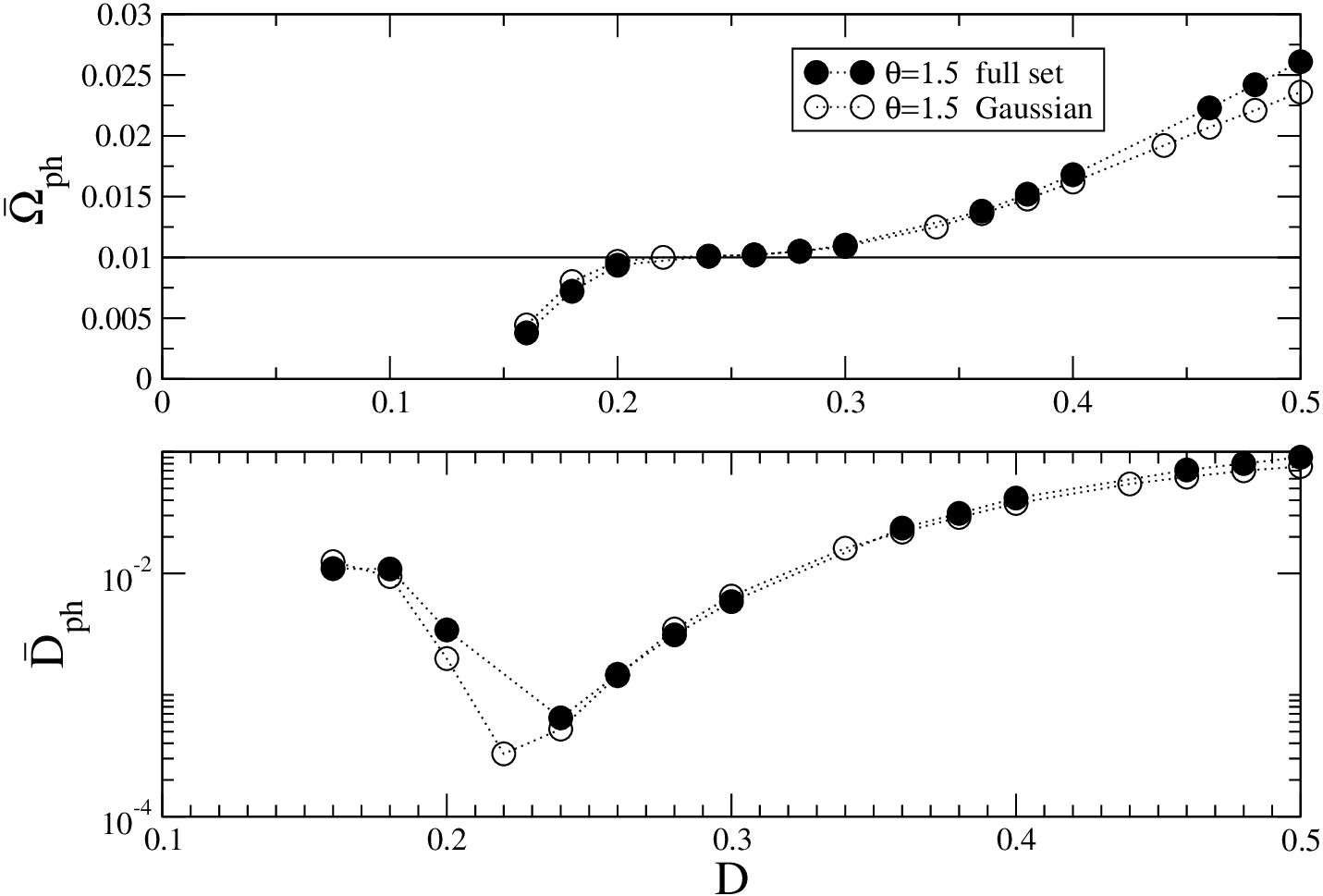}%
\caption{\label{quince} Comparing average phase frequency and phase
  diffusion as obtained from the numerical simulation of the full set of
  equations and from the Gaussian approximation for $N=10$ and
  ($\theta=1.5$). Other parameter values: $\Omega=0.01$
  and $A=0.1$.}
\end{figure}

On the other hand, as depicted in Fig.\ \ref{doce}, the introduction of
even a small coupling between the units drastically changes the
picture. The Gaussian approximation results are in very good agreement
with those obtained from the full set of equations. Indeed, as depicted
in Figs.\ \ref{doce},\ref{trece},\ref{catorce},\ref{quince}, for coupled
units the results of the Gaussian approximation are very good compared
with those obtained from the full set of dynamical equations for some
range of noise values. The range of noise values for phase
synchronization depends on the coupling strength. As $\theta$ is
increased the range decreases. In all cases, synchronization starts
disappearing as the noise strength becomes large. This is to be expected
as large noise values might induce jumps over potential barriers quite
independently of the driving force. Thus, the synchrony between noise
induced jumps and the changes of sign of the external amplitude tends to
disappear. The very same idea of the phase process introduced previously in
Eq.\ (\ref{phase}) loses its meaning for large noise strengths 

\section{\label{conclude}Concluding remarks}
We have analyzed different aspects of the stochastic collective response
of a finite array of globally coupled bistable units to a weak driving
time-periodic force. We focus our analysis on the phenomenon of
stochastic resonance and noise induced phase synchronization. As
demonstrated by our numerical results, both effects might be present in
the collective response as long as the units are not statistically
independent, i. e., their coupling is not zero.

There are two relevant facts: i) the gain of the collective variable
might reach values greater than unity and, ii) the phase frequency
might synchronize with the fundamental driving frequency in wide
ranges of the noise strength. These two features clearly indicate
that, for the weak driving forces considered here, the response of
the system can not be analyzed with the tools of linear response
theory. This is so, even though for a single unit subject to the
same weak driver, LRT provides a very valuable tool to understand SR
at qualitative and even quantitative levels. Our calculations
indicate that the failure of LRT is essentially due to the strong
modification of the output fluctuations brought up by the external
driving and the coupling between the elements of the array. The
output fluctuations become much smaller and short lived than those
present in a single unit system.

We have also analyzed the non-monotonic behavior of the signal-to-noise
ratio with the coupling strength when other parameter values (number
of particles, amplitude and period of the driver) are kept
constant. The largest signal-to-noise ratio occurs at an optimum value
of the coupling strength.

Finally, we have also compared our numerical findings with those
obtained by simplifying approximations that have been put forward in
the literature. In particular, the closure of an infinite hierarchy
of fluctuating cumulant moments at the Gaussian level provides a rather good
description of the simulation results at least for some range of
parameter values. Unfortunately, further simplifications leading to
effective one-dimensional Langevin dynamics for the collective
variable, that seems to provide useful insight on the response for
some regions of parameter space, become invalid in the parameter
region considered in our work.

\begin{acknowledgments}
We acknowledge the support of the Direcci\'on General de Ense\~nanza
Superior of Spain (BFM2005-02884) and the Junta de Andaluc\'{\i}a.
\end{acknowledgments}

\end{document}